\documentclass[preprint, showpacs, superscriptaddress, preprintnumbers, amsmath, amssymb]{revtex4}
\usepackage{graphicx,amssymb,bm,natbib,amsfonts,amsmath,graphics,color}
\usepackage{bm}
\usepackage{hyperref}

\begin{document}

\title{Self-Doping Effects in Epitaxially-Grown Graphene}

\author{D.A. Siegel}
\affiliation{Department of Physics, University of California,
Berkeley, CA 94720, USA}
\affiliation{Materials Sciences Division,
Lawrence Berkeley National Laboratory, Berkeley, CA 94720, USA}

\author{S.Y. Zhou}
\affiliation{Department of Physics, University of California,
Berkeley, CA 94720, USA}
\affiliation{Materials Sciences Division,
Lawrence Berkeley National Laboratory, Berkeley, CA 94720, USA}

\author{F. El Gabaly}
\affiliation{National Center for Electron Microscopy, Lawrence Berkeley National Laboratory, Berkeley, California 94720, USA}

\author{A.V. Fedorov}
\affiliation{Advanced Light Source, Lawrence Berkeley National Laboratory, Berkeley, California 94720, USA}

\author{A. K. Schmid}
\affiliation{National Center for Electron Microscopy, Lawrence Berkeley National Laboratory, Berkeley, California 94720, USA}

\author{A. Lanzara}
\affiliation{Department of Physics, University of California,
Berkeley, CA 94720, USA}
\affiliation{Materials Sciences Division,
Lawrence Berkeley National Laboratory, Berkeley, CA 94720, USA}

\date{\today}

\begin{abstract}
Self-doping in graphene has been studied by examining single-layer epitaxially-grown graphene samples with differing characteristic lateral terrace widths.  Low energy electron microscopy was used to gain real-space information about the graphene surface morphology, which was compared with data obtained by angle-resolved photoemission spectroscopy to study the effect of the monolayer graphene terrace width on the low energy dispersions.  By altering the graphene terrace width we report significant changes in the electronic structure and quasiparticle relaxation time of the material, in addition to a terrace width-dependent doping effect.
\end{abstract}

\maketitle
Single-layer graphene has been the subject of much research in recent years \cite{GeimNatMat} due to the remarkable physics it contains and its potential for device applications.  The electronic properties of graphene sheets are extremely sensitive to their boundary conditions: changing these boundary conditions can create and alter semiconducting band-gaps\cite{Han}, can lead to half-metallicity,\cite{SonNature} and precise control over graphene edge geometries may allow for the creation of complex device components \cite{Nanoletters1}.  Such control is even more important in light of the fact that edge geometries and structural disorder can lead to the formation of impurity bands near the Fermi level,\cite{Peres} effectively hole-doping the system and substantially altering the transport properties of graphene.  Although the effect of this self-doping in ribbons has been analyzed theoretically,\cite{Peres} so far no systematic experimental study exists.  Transport studies of graphene nanoribbons have been performed, \cite{Han, XWang} but the p-doping may be attributed to the patterning process and atmospheric contaminants.  Similarly, although evidence of self doping can be found in Raman measurements,\cite{Ferrari1,Ferrari2,Casiraghi} the doping levels vary greatly among these samples even in the absence of measured disorder.
The primary difficulty is in the comparison of the microscopic features of disorder to spectroscopy data: the size of the probing beam is often too large to study the position-dependence of the self-doping as a function of the distance from a sample edge.
Instead, controlling the characteristic length scale of the sample itself can be a more practical solution.  This can be achieved by growing epitaxial graphene on 6H-SiC, as it forms a tangle of broad nanoribbon-like monolayer graphene terraces\cite{OhtaLEEM, Hibino}, whose size can be directly monitored by low energy electron microscopy (LEEM)\cite{Bauer} during the growth process.  The effect of disorder and/or self-doping on the electronic structure of graphene can then be directly studied on the same samples by angle-resolved photoemission spectroscopy (ARPES)\cite{damascelli}.

In this letter we present a systematic study of the self-doping on epitaxially-grown graphene through the edges of the graphene terraces.  Several epitaxially-grown graphene samples with average monolayer terrace widths ranging from 50nm to 180nm have been studied by LEEM and ARPES.  As the terrace widths increase we observe an increase in charge carrier density, resulting in a rigid shift of the low binding-energy bandstructure and a constant Fermi velocity, and a corresponding decrease of the quasiparticle lifetime due to the increase in disorder.  These results provide direct experimental evidence of self-doping effects in graphene and illuminate the role of finite graphene ribbon widths on its electronic structure.

Samples were grown on the Si-terminated face of 6H-SiC substrate as previously reported \cite{Rollings}.  LEEM was performed at the National Center for Electron Microscopy at Lawrence Berkeley National Laboratory\cite{OhtaLEEM,Hibino}.  High resolution ARPES data were taken at BL12.0.1 of the Advanced Light Source at a temperature of 15K with 50eV photons after annealling samples to 1000C.

Figures 1(a-c) show LEEM images of three characteristic epitaxially-grown graphene samples.\cite{OhtaLEEM,Hibino}  The width of the single-layer graphene terraces has been extracted using a standard method \cite{Wang}, by first identifying outlines of the single-layer terraces, then drawing straight lines in the images, and finally measuring the linear widths of the terraces crossed by the lines.  The terrace size increases from 50nm to 180nm from panel a to panel c.

In figure 2, the raw ARPES intensity maps (lower panels) and area-normalized momentum distribution curves (MDCs), intensity profiles at constant energy (upper panels), are shown for five samples, S1-S5, ordered by increasing terrace width.  Data are taken along a high symmetry direction through the K point (see horizontal black line in the inset of panel a).  As the terrace widths increase the following qualitative trends are observed (quantitative discussion follows in figures 3 and 4): 1) The separation between the two branches that disperse towards E$_F$ (dotted line) increases (compare the separation between the MDCs peaks in the upper panel); 2) The intensity maps become sharper at the Fermi level and higher binding energy (compare width of the MDCs peak in the upper panel); 3) The region of vertical intensity separating the conduction band from the valence band, namely the gap,\cite{ZhouGap} decreases in agreement with a previous report.\cite{ZhouReply}  It can also be seen in panels d and e that additional bands appear faintly (see bottom of the conduction band and a replica of the valence band) due to an increase in the ratio of the bilayer to the monolayer content as its coverage increases.

A quantitative analysis of the Fermi level MDCs is presented in Figure 3.  For ease of comparison, the MDCs curves shown in the previous figure have been offset and normalized by the area underneath, after removing a constant background (see panel a). The direct comparison clearly shows a shift of the MDC peak positions toward higher momentum and a decrease of the MDC widths with increasing terrace size, pointing to a change of the doping level and scattering rate, respectively.
The charge density (panel b) and quasiparticle scattering rate (panel c) as a function of the monolayer graphene terrace widths are shown.  
The charge density, a quantity directly related to the separation of the MDCs peak at $E_F$, is calculated by assuming the electron pockets of graphene to be circles in k-space\cite{Slonczewski} with diameters given by the separations of the dispersions at the Fermi level, and invoking Luttinger's theorem to extract the charge carrier density from the Fermi surface area.  Clearly the size of the electron pocket \textit{decreases}, or the hole doping \textit{increases}, as the monolayer graphene terrace width \textit{increases} from 50nm to 180nm, in line with the shift of the MDC peak position of panel a.  This is in agreement with the prediction that self-doping varies inversely with the graphene ribbon width\cite{Peres}.  More specifically, the data can be well described by the function $\sigma = c_{1} W^{-1} + c_{2}$ (gray dashed line in panel b), where $\sigma$ is the charge density, W is the terrace width, and $c_{1}$ and $c_{2}$ are constants.  We find that the constant $c_{1}$=-2$\times$10$^7$ cm$^{-1}$ is very close to the approximate theoretical value of -5$\times$10$^6$ cm$^{-1}$ (using t'=0.1eV, and a dielectric constant of 10 in Eq. 161 of ref \cite{Peres}).  In the limit of an infinite epitaxially-grown graphene plane, one would expect an electron-doping of approximately $c_{2}$ = 1.17$\times$10$^{13}$ cm$^{-2}$ due to the interaction with the substrate.

Despite the overall agreement between theory and experiment, the theoretical prediction for the change in doping as the ribbon width changes from 34nm to 170nm is 6.5$\times$10$^{11}$ electrons/cm$^2$, which is a factor of $\sim$4 smaller than the experimentally observed doping for similar widths.  There are several reasons why this might be expected.   First, the method used for extracting the terrace widths from LEEM images is likely an overestimate of the actual ribbon width, since an arbitrary line drawn across a LEEM image will not cross the ribbon edges perpendicularly, and because the resolution of the LEEM is finite.  Second, the experimental geometry is different from the theoretical one: jagged edges might permit more self-doping than straight ones, and higher densities of other types of lattice defects may be associated with real ribbon edges.  Finally, the details of the self-doping of epitaxially-grown graphene might be different from the free-standing case.

Figure 3(c) shows the Fermi level MDC width, a quantity directly related to the inverse quasiparticle relaxation time, $\Gamma$ $\sim$ $1/\tau$, as a function of the monolayer graphene terraces width.  
As the terrace widths increase we observe a clear decrease of the MDC width, or increase of the quasiparticle scattering rate, in line with panel (a) (compare horizontal arrows).
If the quasiparticle scattering rate were dominated by the terrace edges, the mean free path, and therefore the transport lifetime, would vary linearly with the terrace width, although it should be noted that the single-particle relaxation time probed by ARPES may not equal the transport lifetime\cite{Hwang}.  Therefore, as a guide to the eye, a dotted line is plotted of the form $FWHM = c_{3} W^{-1} + c_{4}$, where FWHM is the MDC width, W is the terrace width, and $c_{3}$ and $c_{4}$ are constants equal to 7.5 and 0.0125$\AA$$^{-1}$, respectively.  This change in lifetime reflects the decrease of the quasiparticle mean free path with decreasing monolayer graphene terrace width and provides further confirmation of the observation of self-doping in graphene.

The effect of finite terrace size on the conical dispersion is shown in figure 4.  The energy vs momentum dispersion relation are obtained by fitting the MDCs spectra over the full energy range, using the standard Lorentzian fitting procedure. \cite{damascelli}  
Dispersions for binding energies lower than the Dirac point ($|E - E_F| < |E - E_D|$) are displayed in panel a, while dispersions at higher binding energy ($|E - E_F| > |E - E_D|$) are displayed in panel b.  With increasing terrace width the data show a clear rigid shift of the dispersion away from the K point at low binding energy (in line with the observed shift of the MDC peaks at k$_F$ in panel a) and a rigid shift of the dispersions toward the K point at high binding energy. These shifts, together with the nearly constant Fermi velocity (panel c) extracted from the low energy dispersions v$_F$=$\frac{1}{\hbar}\frac{\partial{E}}{\partial{k}}$, point to a rigid shift of the conduction band in energy, and consequently an increase of the chemical potential, as the terrace widths increase.  For a schematic summary see inset of panel a.  Interestingly, although on a qualitative level the shift observed in the high energy dispersion of panel b is also in agreement with this picture, on a quantitative level the magnitude of the shift is smaller by roughly a factor of 2.  This apparent discrepancy can be partially accounted for by the reported change in the size of the gap at the Dirac point \cite{ZhouReply}, but further investigation is required to exactly determine the mechanism behind this change.

In conclusion, we have manipulated the electronic structure of monolayer graphene by adjusting the widths of the monolayer terraces during the growth process of epitaxially grown graphene.  We found that altering the graphene terrace width has a significant effect on the electronic structure of the material, even for samples with length scales on the order of 100nm.  These finite-size effects will play a role in electronics applications as graphene-based components of various sizes and geometries are fabricated in the future.

\begin{acknowledgments}
We thank D.H. Lee, A. Castro Neto and J. Graf for useful discussions. This work was supported by the Director, Office of Science, Office of Basic Energy Sciences, Materials Sciences and Engineering Division, of the U.S. Department of Energy under Contract No. DE-AC02-05CH11231, and by the National Science Foundation through Grant No.~DMR03-49361.
\end{acknowledgments}


\pagebreak

\pagebreak
\noindent
Figure captions:\\

\noindent
Fig. 1: (Color online) (a)-(c) 2$\mu$m $\times$ 2$\mu$m LEEM images of three samples, with red dots to denote the typical terrace widths used in this paper.  The three shades in these images, white, grey, and black, are buffer layer, monolayer graphene, and bilayer graphene, respectively.  Inset of (a): Reflectivity as a function of incident kinetic energy for the buffer layer and first monolayer through fourth monolayer (bottom to top).  The number of dips in the reflectivity curves is equal to the number of graphene layers.  The dotted line denotes 4.2eV electron kinetic energy, which is used in (a)-(c).

\noindent
Fig. 2: (Color online) (a-e) ARPES spectra of samples with increasing monolayer graphene terrace width, S1 through S5. Inset of (a) shows the Brillouin zone of graphene.  The horizontal line through the K point shows the measurement geometry in k-space.  The large and small pink left-right arrows give the peak separation and width of the Fermi level MDC of S1.

\noindent
Fig. 3: (Color online) (a) The Fermi level MDCs with incremental offset.  (b) The electronic charge carrier density as a function of terrace width.  (c) Fermi level MDC widths, extracted from panel a.

\noindent
Fig. 4: (Color online) Low (a) and high (b) binding energy dispersions as a function of terraces widths.  Note that the small curvature at E$_F$ in panel a is due to the finite energy resolution of the experiment.  (c) Fermi velocities vs terrace widths.  Inset of (a, b) shows a cartoon of the graphene band structure as a function of terraces width.  The boxed regions in the insets denote the regions of the dispersions of Figs. 4(a) and 4(b).

\pagebreak
\begin{figure}[p]
\caption{\label{Figure 1}}
\includegraphics[width=8.5 cm] {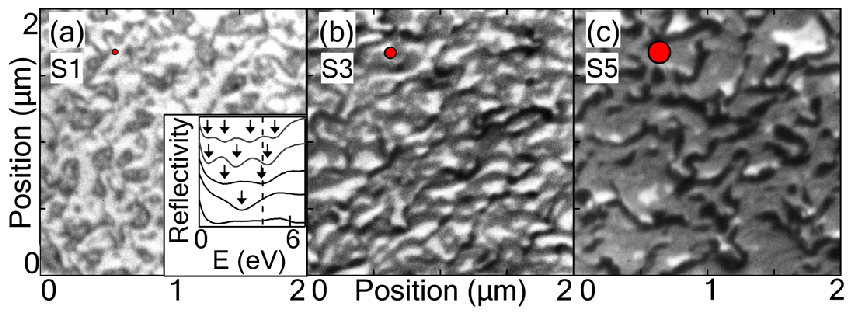}
\end{figure}

\begin{figure} [p]
\caption{\label{Figure 2}}
\includegraphics[width=8.5 cm] {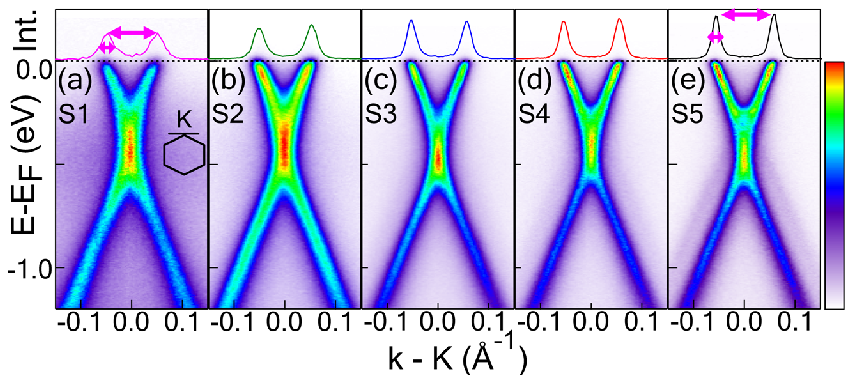}
\end{figure}

\begin{figure} [p]
\caption{\label{Figure 3}}
\includegraphics[width=8.5 cm] {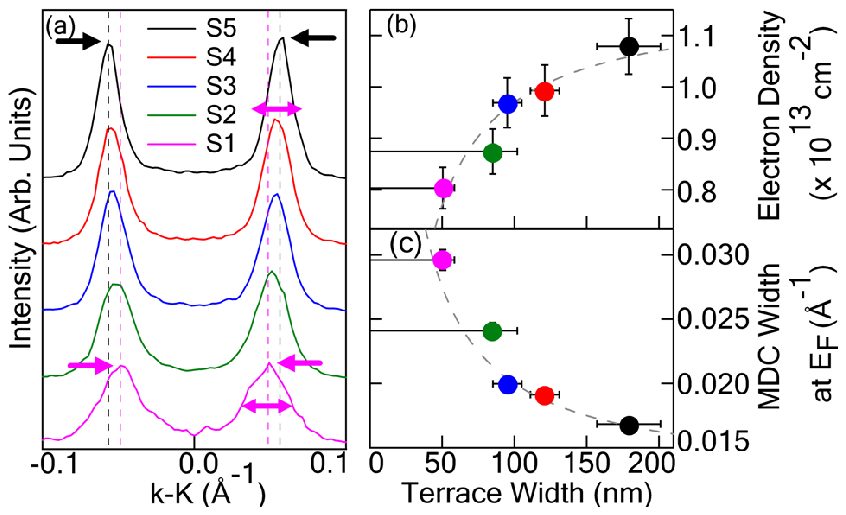}
\end{figure}

\begin{figure} [p]
\caption{\label{Figure 4}}
\includegraphics[width=8.5 cm] {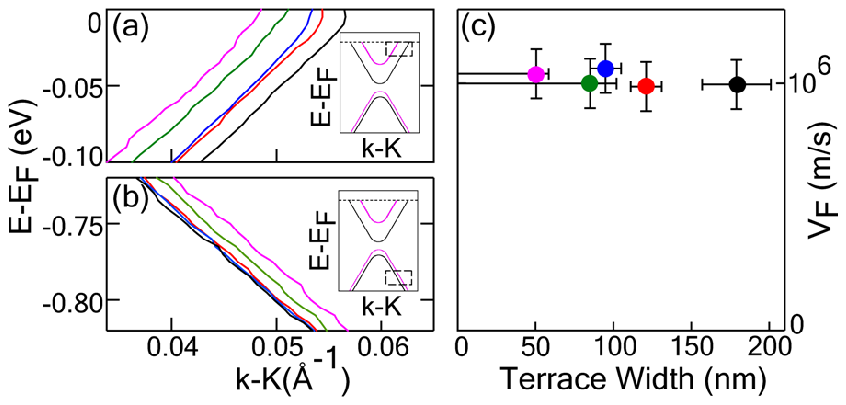}
\end{figure}

\end{document}